\documentclass{revtex4}
\usepackage{amsthm,amssymb,amsmath}				
\usepackage{graphicx,comment}
\usepackage{float}   
 \usepackage{xcolor}

\begin{document}

\title{On the Klein-Gordon oscillators in Eddington-inspired Born-Infeld gravity global monopole spacetime and a Wu-Yang magnetic monopole}
\author{Omar Mustafa}
\email{omar.mustafa@emu.edu.tr}
\affiliation{Department of Physics, Eastern Mediterranean University, 99628, G. Magusa, north
Cyprus, Mersin 10 - Turkiye.}

\author{A. R. Soares}
\email{adriano.soares@ifma.edu.br}
\affiliation{Instituto Federal de Educa\c{c}\~ao Ci\^encia e Tecnologia do Maranh\~ao,  R. Dep. Gast\~ao Vieira, 1000, CEP 65393-000 Buriticupu, MA, Brazil.}

\author{C. F. S. Pereira} 
\email{carlos.f.pereira@edu.ufes.br}
\affiliation{Departamento de F\'isica e Qu\'imica, Universidade Federal do Esp\'irito Santo, Av.Fernando Ferrari, 514, Goiabeiras, Vit\'oria, ES 29060-900, Brazil.}

\author{R. L. L. Vit\'oria} 
\email{ricardo-luis91@hotmail.com}
\affiliation{Faculdade de F\'isica, Universidade Federal do Par\'a, Av. Augusto Corr\^ea, Guam\'a, 66075-110, Bel\'em, PA, Brazil.}

\begin{abstract}
\textbf{Abstract:}\ We consider Klein-Gordon (KG) particles in a global monopole (GM) spacetime within Eddington-inspired Born-Infeld gravity (EiBI-gravity) and in a Wu-Yang magnetic monopole (WYMM).  We discuss a set of KG-oscillators in such spacetime settings. We propose a textbook power series expansion for the KG radial wave function that allows us to retrieve the exact energy levels for KG-oscillators in a GM spacetime and a WYMM without EiBI-gravity. We, moreover, report some \textit{conditionally exact}, closed form, energy levels (through some parametric correlations) for KG-oscillators in a GM spacetime and a WYMM within EiBI-gravity, and for massless KG-oscillators in a GM spacetime and a WYMM within EiBI-gravity under the influence of a Coulomb plus linear Lorentz scalar potential. We study and discuss the effects of the Eddington parameter $\kappa$, GM-parameter $\alpha$, WYMM strength $\sigma$, KG-oscillators' frequency $\Omega$, and the coupling parameters of the Coulomb plus linear Lorentz scalar potential, on the spectroscopic structure of the KG-oscillators at hand. Such effects are studied over a vast range of the radial quantum number $n_r\geq 0$ and include energy levels clustering at  $\kappa>>1$ (i.e., extreme EiBI-gravity), and at  $|\sigma|>>1$ (i.e., extreme WYMM strength).

\textbf{PACS }numbers\textbf{: }05.45.-a, 03.50.Kk, 03.65.-w

\textbf{Keywords:} Klein-Gordon oscillators, Global monopole spacetime, Wu-Yang magnetic monopole, Eddington-inspired Born-Infeld gravity.
\end{abstract}

\maketitle

\section{Introduction}

Among the well known topological defects in the primordial universe, due to
phase transition during the rapid expansion and cooling process \cite%
{Re1,Re2}, are domain walls \cite{Re020,Re021}, cosmic strings \cite%
{Re021,Re3,Re4,Re5,Re6} and global monopoles (GM) \cite{Re2,Re3,Re6,Re7}.
Global monopoles are characterized by spontaneous global symmetry breaking
and behave like elementary particles, with their energy is mostly
concentrated near the monopole core \cite{Re7}. Global monopoles are objects
that modify the geometry of spacetime \cite{Re3,Re6,Re7}, and are
spherically symmetric topological defects with the line element%
\begin{equation}
ds^{2}=-B\left( r\right) \,dt^{2}+A\left( r\right) \,dr^{2}+r^{2}\left(
d\theta ^{2}+\sin ^{2}\theta \,d\varphi ^{2}\right) .  \label{eq1}
\end{equation}%
In their study of gravitational field of a GM, Barriola and Vilenkin (BV) 
\cite{Re2} have shown that the monopole, effectively, exerts no
gravitational force, and the space around and outside the monopole has a
solid deficit angle that deflects all light. They have shown that%
\begin{equation}
B\left( r\right) =A\left( r\right) ^{-1}=1-8\pi G\eta ^{2}-\frac{2GM}{r},
\label{eq2}
\end{equation}%
where $M$ is a constant of integration and in the flat space $M\sim M_{core}$
( $M_{core}$ is the mass of the monopole core). \ Assuming very light GMs,
one may neglect the mass term and rescale the variables $r$ and $t$ \cite%
{Re7}, so that the GM metric reads 
\begin{equation}
ds^{2}=-dt^{2}+\frac{1}{\alpha ^{2}}dr^{2}+r^{2}\left( d\theta ^{2}+\sin
^{2}\theta \,d\varphi ^{2}\right) ,  \label{e01}
\end{equation}%
where $0<\alpha ^{2}=1-\beta \leq 1$, where $\beta =8\pi G\eta ^{2}$ is the
deficit angle, $\alpha $ is a global monopole parameter that depends on the
energy scale $\eta $, and $G$ is the gravitational constant \cite%
{Re7,Re8,Re9,Re91,Re10}. Obviously, this metric collapses into the flat
Minkowski one when $\alpha =1$.

The BV-solution in \cite{Re2} has extensively inspired research studies.
Within the $f\left( \mathbf{R}\right) $ theories of gravity, for example,
spacetime geometry around GM is studied \cite{Re100}, vacuum polarization
effects in the presence of a Wu-Yang\ \ magnetic monopole (WYMM) \cite%
{Re101,Re1011,Re102}, gravitating magnetic monopole \cite{Re103}, The
effects of GM spacetime background on the quantum mechanical spectroscopic
structures are studied, for both relativistic and non-relativistic. To
mention a few, Dirac and Klein-Gordon (KG) oscillators \cite{Re11}, Schr\"{o}%
dinger oscillators \cite{Re9}, Schr\"{o}dinger oscillators in a GM spacetime
and a Wu-Yang magnetic monopole \cite{Re91}, KG particles with a dyon,
magnetic flux and scalar potential \cite{Re8}, bosons in Aharonov-Bohm flux
field and a Coulomb potential \cite{Re131}, Schr\"{o}dinger particles in a
Kratzer potential \cite{Re132}, Schr\"{o}dinger particles in a Hulth\.{e}n
potential \cite{Re1321} , scattering by a monopole \cite{Re133}, Schr\"{o}%
dinger particles in a Hulth\.{e}n plus Kratzer potential \cite{Re1331},
KG-oscillators, AB-effect \cite{Re1332}, quark-antiquark interaction \cite{me}, thermodynamical properties of a quantum particle
confined into two elastic concentric spheres \cite{me1} and in small oscillations on a diatomic molecule \cite{me2}. Yet, the influence of
topological defects associated with different spacetime backgrounds on the
spectroscopic structure of quantum mechanical systems have been a subject of
research attention over the years. Like, Dirac and Klein-Gordon (KG)
oscillators are studied in a variety spacetime structures, e.g., \cite%
{Re11,Re12,Re121,Re13,Re14,Re15,Re16,Re17,Re18,Re19,Re20,Re21,Re211,Re212,Re213,Re22,Re23,Re24,Re25,Re26,Re27,Re271,Re272}%
.

Very recently, the incorporation of the nonlinear electrodynamics of
Born-Infeld into Eddington theory of gravity \cite{ref1,ref11,ref12} has
introduced the so called Eddington-inspired Born-Infeld (EiBI) theory of
gravity (EiBI-gravity, in short). While the EiBI theory is known to be
equivalent to the Einstein's General Relativity (GR) in vacuum, it provides
distinctive features when matter is included, and it possesses internal
consistency for being free of instabilities and ghosts \cite{ref2}. One of
the most interesting feature of the EiBI-gravity is its ability to avoid
cosmological singularity and yields entirely singularity-free states \cite%
{ref1,ref3}. It has been reported by \cite{ref3,ref4,ref5,ref6} (to mention
a recent few) that the GM spacetime in EiBI-gravity, generated by a source
matter, is described, in spherical coordinates, by the metric%
\begin{equation}
ds^{2}=-\alpha ^{2}\,dt^{2}+\frac{r^{2}}{\alpha ^{2}\left( r^{2}+\kappa
\beta \right) }\,dr^{2}+r^{2}\,d\Omega ^{2},  \label{eq.1}
\end{equation}%
where $\kappa $ is the Eddington parameter that controls the nonlinearity of
the EiBI-gravity, and $d\Omega ^{2}=d\theta ^{2}+\sin ^{2}\theta \,d\varphi
^{2}$. It is obvious that this metric reduces into the GM one, Eq.(\ref{e01}%
) above, when the Eddington parameter is switched off, ie.,  $\kappa =0$,
hence it is occasionally called modified GM or EiBI-GM. The study of the
effects of the gravitational field, manifestly introduced by the GM-spacetime
in EiBI-gravity, on the quantum mechanical spectroscopic structure were only
very recently carried out by Pereira et al \cite{ref5,ref7}. Pereira and
co-authors \cite{ref5,ref7} were able to discuss one state (the lowest state
with $n=1$ for any value of the angular momentum quantum number $\ell $).
Whilst we emphasis that the the procedure and methodology they have used is
just fine and good, their methodology does not work for $n>1$. In the
current methodical proposal, however, we report/provide an alternative
approach to obtain a \emph{conditionally exact solution} (or if you wish, 
\emph{quasi-exact solution}) \cite{ref8} for the Klein-Gordon (KG)
oscillator in a GM-spacetime within EiBI-gravity. Namely, we shall discuss
KG-oscillators in a GM-spacetime and a Wu-Yang magnetic monopole (WYMM)  
\cite{Re101,Re1011,Re102} within EiBI-gravity for any $n$ and $\ell $ (at a
very specific correlation/constraint between the KG-oscillator's frequency
and Eddington parameter). 

In so doing, we are motivated by the fact that the influence of the
topological defects, generated by gravitational fields, on the quantum
mechanical systems, are interesting not only for quantum gravity but also
for the geometrical theory of topological defects in condensed matter
physics (e.g., \cite{ref9,ref10,ref011}). To the best of our knowledge, the
studies of Pereira and co-authors \cite{ref5,ref7} are the only attempts
made, in the literature, to study the KG-oscillators in a GM-spacetime
within EiBI-gravity. In the current methodical proposal we shall add (in
addition to Pereira et al.'s \cite{ref5,ref7} approach) yet another degree of
freedom that allows the reader to observe the effects of EiBI-GM
gravitational field on the full spectrum of the KG-oscillators in a WYMM.
Nevertheless, one should be reminded that although the WYMM is a
theoretical concept that is yet to be observed, it remains an interesting
model for modern theoretical particle physics and topological properties in
gauge theories. 

The organization of the current proposal is in order. In section 2, we
consider KG-particles in GM spacetime within EiBI-gravity and in a WYMM, and
recollect/recycle the corresponding angular and radial parts of the
KG-equations. In section 3, we discuss KG-oscillators in a GM spacetime
within EiBI-gravity and in a WYMM. Therein, we argue that although the
solution of the corresponding Schr\"{o}dinger-like KG radial equation has a
solution in the form of confluent Heun functions $H_{C}\left( a,b,c,d,f,%
\frac{r^{2}+\kappa }{\kappa }\right) $, the power-series expansion should
not be dictated by the argument of the confluent Heun functions (i.e., $%
\left( r^{2}+\kappa \right) /\kappa $) but rather a textbook one that
allows some freedom in the manipulations of a three recursion relations
(usual product of such Heun equations) that in turn allow feasible reduction of
such recursion relation into that of the KG-oscillator in no 
Eddington gravity (i.e. $\kappa =0$). We show, in the same section,
how to obtain the energy levels for the KG-oscillators in a GM spacetime in
a Wu-Yang magnetic monopole in no EiBI-gravity, $\tilde{\kappa}=0$, from the
same recursion relations, in subsection 3-A. In subsection 3-B, we report a
conditionally exact solution, through a parametric correlation, and obtain
some conditionally exact energy levels for the KG-oscillators in a GM
spacetime within EiBI-gravity, $\tilde{\kappa}\neq 0$, and a WYMM. Moreover,
we discuss, in subsection 3-C, the conditionally exact energy levels for
some massless KG-oscillators in a GM spacetime within EiBI-gravity and in a
WYMM under the influence of a Coulomb plus linear Lorentz scalar potential.
Our concluding remarks are given in section 4.

\section{KG-particles in GM spacetime within E\lowercase{i}BI-gravity and in a WYMM}

A rescaling in the forms of $\sqrt{\left( 1-\beta \right) }\,dt\rightarrow dt
$ along with $\tilde{\kappa}=\kappa \beta $, would allow us to rewrite
metric (\ref{eq.1}) as 
\begin{equation}
ds^{2}=-\,dt^{2}+\frac{1}{\alpha ^{2}\left( 1+\frac{\tilde{\kappa}}{r^{2}}%
\right) }\,dr^{2}+r^{2}\,d\Omega ^{2}.  \label{eq.2}
\end{equation}%
It should be noted that $\tilde{\kappa}<0$ would describe a topologically
charged wormhole \cite{ref012,ref013,ref014}, $\alpha =1$ and $\tilde{\kappa}%
<0$ corresponds to Morris-Thorne-type wormhole spacetime \cite{ref015,ref016}%
, $\kappa =0$ would describe a GM spacetime, and $\kappa >0$ corresponds to
a GM spacetime in EiBI-gravity. The corresponding metric tensor is 
\begin{equation}
g_{\mu \nu }=\left( 
\begin{tabular}{cccc}
$-1$ & $0$ & $0$ & $0$ \\ 
$0$ & $\;\alpha ^{-2}\left( 1+\frac{\tilde{\kappa}}{r^{2}}\right) ^{-1}$ & $0
$ & $0\medskip $ \\ 
$0$ & $0$ & $r^{2}$ & $0\medskip $ \\ 
$0$ & $0$ & $0$ & $r^{2}\sin ^{2}\theta $%
\end{tabular}%
\right) ;\;\mu ,\nu =t,r,\theta ,\varphi ,  \label{eq.3}
\end{equation}%
to imply 
\begin{equation}
\det \left( g_{\mu \nu }\right) =g=-\frac{r^{4}\sin ^{2}\theta }{\alpha
^{2}\left( 1+\frac{\tilde{\kappa}}{r^{2}}\right) },  \label{eq.4}
\end{equation}%
and 
\begin{equation}
g^{\mu \nu }=\left( 
\begin{tabular}{cccc}
$-1$ & $0$ & $0$ & $0$ \\ 
$0$ & $\;\alpha ^{2}\left( 1+\frac{\tilde{\kappa}}{r^{2}}\right) $ & $0$ & $%
0\medskip $ \\ 
$0$ & $0$ & $1/r^{2}$ & $0\medskip $ \\ 
$0$ & $0$ & $0$ & $1/r^{2}\sin ^{2}\theta $%
\end{tabular}%
\right) .  \label{eq.5}
\end{equation}%
Then, the KG-equation would reads 
\begin{equation}
\left( \frac{1}{\sqrt{-g}}\tilde{D}_{\mu }\sqrt{-g}g^{\mu \nu }\tilde{D}%
_{\nu }\right) \,\Psi \left( t,r,\theta ,\varphi \right) =\left[ m_{\circ
}+S\left( r\right) \right] ^{2}\,\Psi \left( t,r,\theta ,\varphi \right) ,
\label{eq.6}
\end{equation}%
where $\tilde{D}_{\mu }=D_{\mu }+\mathcal{F}_{\mu }$ is in a non-minimal
coupling form with $\mathcal{F}_{\mu }$ $\in 
\mathbb{R}
$, $D_{\mu }=\partial _{\mu }-ieA_{\mu }$ is the gauge-covariant derivative
that admits minimal coupling, $A_{\nu }=\left( 0,0,0,A_{\varphi }\right) $
is the 4-vector potential, $S\left( r\right) $ is the the Lorentz scalar
potential, and $m_{\circ }$ is the rest mass energy (i.e., $m_{\circ }\equiv
m_{\circ }c^{2}$, with $\hbar =c=1$ units to be used through out this
study). Moreover, we use $\mathcal{F}_{\mu }=\left( 0,\mathcal{F}%
_{r},0,0\right) $ to incorporate the KG-oscillators in the GM spacetime in
EiBI-gravity. In particular, we are considering KG-oscillators \cite{kgo}, as it is a relativistic quantum model that has been extensively studied and applied in the literature, due to its analyticity and the possibility of recovering the Shr\"odinger oscillator in the non-relativistic limit \cite{kgo1}. It is worth mentioning that the oscillator goes beyond a mere academic exercise, as any central potential, which has a minimum point, around this point, it behaves like an oscillator \cite{me3}. Furthermore, oscillations are found in various systems from varied areas of physics \cite{eisb}. Consequently, the KG-equation (\ref{eq.6}) would read%
\begin{gather}
\left\{ -\partial _{t}^{2}+\frac{\alpha ^{2}}{r^{2}}\sqrt{1+\frac{\tilde{%
\kappa}}{r^{2}}}\,\left( \partial _{r}+\mathcal{F}_{r}\right) \,r^{2}\,\sqrt{%
1+\frac{\tilde{\kappa}}{r^{2}}}\left( \partial _{r}-\mathcal{F}_{r}\right) +%
\frac{1}{r^{2}}\left[ \frac{1}{\sin \theta }\,\partial _{\theta }\,\sin
\theta \,\partial _{\theta }\right. \right. \medskip   \notag \\
\left. \left. +\frac{1}{\sin ^{2}\theta }\left( \partial _{\varphi
}-ieA_{\varphi }\right) ^{2}\,\right] \right\} \Psi \left( t,r,\theta
,\varphi \right) =\left[ m_{\circ }+S\left( r\right) \right] ^{2}\,\Psi
\left( t,r,\theta ,\varphi \right) .  \label{eq.7}
\end{gather}%
With the substitution of 
\begin{equation*}
\Psi \left( t,r,\theta ,\varphi \right) =\Psi \left( t,\rho ,\theta ,\varphi
\right) =e^{-iEt}\phi \left( r\right) Y_{\sigma ,\ell ,m}\left( \theta
,\varphi \right) ,
\end{equation*}%
we obtain%
\begin{equation}
\left\{ E^{2}+\frac{\alpha ^{2}}{r^{2}}\sqrt{1+\frac{\tilde{\kappa}}{r^{2}}}%
\,\left( \partial _{r}+\mathcal{F}_{r}\right) \,r^{2}\,\sqrt{1+\frac{\tilde{%
\kappa}}{r^{2}}}\left( \partial _{r}-\mathcal{F}_{r}\right) -\frac{\lambda }{%
r^{2}}\right\} \phi \left( r\right) =\left[ m_{\circ }+S\left( r\right) %
\right] ^{2}\phi \left( r\right) ,  \label{eq.8}
\end{equation}%
where%
\begin{equation}
\left[ \frac{1}{\sin \theta }\,\partial _{\theta }\,\sin \theta \,\partial
_{\theta }+\frac{1}{\sin ^{2}\theta }\left( \partial _{\varphi
}-ieA_{\varphi }\right) ^{2}\,\right] Y_{\sigma ,\ell ,m}\left( \theta
,\varphi \right) =-\lambda Y_{\sigma ,\ell ,m}\left( \theta ,\varphi \right)
.  \label{eq.9}
\end{equation}%
At this point, one should notice that $\lambda =\ell \left( \ell +1\right) $
for $A_{\varphi }=0$ case, and for the WYMM \cite{Re101,Re1011,Re102} $%
\lambda $ is given by 
\begin{equation}
\lambda =\ell \left( \ell +1\right) -\sigma ^{2},  \label{eq.10}
\end{equation}%
where $\sigma =eg$ (with $g$ is the WYMM strength). The details of the
result in Eq.(\ref{eq.10}) are given in the Appendix (to keep this study
self-contained). Consequently, Eq.(\ref{eq.8}) would simply read%
\begin{equation}
\left\{ E^{2}+\frac{\alpha ^{2}}{r^{2}}\sqrt{1+\frac{\tilde{\kappa}}{r^{2}}}%
\,\left( \partial _{r}+\mathcal{F}_{r}\right) \,r^{2}\,\sqrt{1+\frac{\tilde{%
\kappa}}{r^{2}}}\left( \partial _{r}-\mathcal{F}_{r}\right) -\frac{\lambda }{%
r^{2}}\right\} \phi \left( r\right) =\left[ m_{\circ }+S\left( r\right) %
\right] ^{2}\phi \left( r\right) .  \label{eq.11}
\end{equation}%
With $\mathcal{F}_{r}=\Omega r$, $S\left( r\right) =0$, and $\sigma =0$ (no
WYMM), this equation would imply%
\begin{equation}
\left\{ \left( 1+\frac{\tilde{\kappa}}{r^{2}}\right) \partial
_{r}^{2}\,+\left( \frac{2}{r}+\frac{\tilde{\kappa}}{r^{3}}\right) \partial
_{r}+\left[ K_{1}^{2}-\frac{K_{2}^{2}}{r^{2}}-K_{3}^{2}\,r^{2}\right]
\right\} \phi \left( r\right) =0,  \label{eq.12}
\end{equation}%
where%
\begin{equation}
K_{1}^{2}=\frac{E^{2}-m_{\circ }^{2}}{\alpha ^{2}}-3\Omega -\Omega ^{2}%
\tilde{\kappa},\;K_{2}^{2}=\frac{\ell \left( \ell +1\right) }{\alpha ^{2}}%
+2\Omega \,\tilde{\kappa},\;K_{3}^{2}=\Omega ^{2}.  \label{eq.13}
\end{equation}%
This result is in exact accord with equations (6) and (7) reported by
Pereira et al. \cite{ref5} (where our $\Omega =m\omega $ of Pereira et al. 
\cite{ref5}). In the following section, we shall consider the inclusion of a
WYMM and describe a different solution to the problem at hand (i.e., a
conditionally exact solution that deals not only with $n=1$ states but also
with $n\geq 1$ states).

\section{KG-oscillators in a GM spacetime within E\lowercase{i}BI-gravity and in a WYMM}

In what follows we shall consider the more general case that includes the
WYMM and rewrite equation (\ref{eq.11}) with $\mathcal{F}_{r}=\Omega r$, $%
S\left( r\right) =0$, and $\sigma \neq 0$ so that%
\begin{equation}
\left\{ \left( 1+\frac{\tilde{\kappa}}{r^{2}}\right) \partial
_{r}^{2}\,+\left( \frac{2}{r}+\frac{\tilde{\kappa}}{r^{3}}\right) \partial
_{r}-\Omega ^{2}r^{2}-\frac{\tilde{\ell}\left( \tilde{\ell}+1\right) }{r^{2}}%
+\mathcal{E}^{2}\right\} \phi \left( r\right) =0.  \label{eq.14}
\end{equation}%
This equation represents KG-oscillators in GM spacetime in EiBI-gravity and
a WYMM, where%
\begin{equation}
\mathcal{E}^{2}=\frac{E^{2}-m_{\circ }^{2}}{\alpha ^{2}}-3\Omega -\Omega ^{2}%
\tilde{\kappa},\;\tilde{\ell}\left( \tilde{\ell}+1\right) =\frac{\ell \left(
\ell +1\right) -\sigma ^{2}}{\alpha ^{2}}+2\Omega \,\tilde{\kappa}.
\label{eq.15}
\end{equation}%
and hence 
\begin{equation}
\tilde{\ell}=-\frac{1}{2}+\sqrt{\frac{1}{4}+\frac{\ell \left( \ell +1\right)
-\sigma ^{2}}{\alpha ^{2}}-2\Omega \,\tilde{\kappa}}.  \label{q.16}
\end{equation}%
$\tilde{\ell}$ \ is a new irrational angular momentum quantum number that
collapses into the regular one $\ell =0,1,2,\cdots $ for $\alpha =1$ (i.e.,
flat Minkowski spacetime)$,\sigma =0$ (no Wu-Yang monopole), and $\tilde{%
\kappa}=0$ (i.e., no EiBI-gravity). One should also notice that the case of $%
\tilde{\kappa}=0,$ $\alpha \neq 1$, and $\sigma \neq 0$ corresponds to the
GM spacetime background with a WYMM. This fact should provide a controling
mechanism on the validity of the solution of the more general case that
includes EiBI-gravity. 

Apriori, one should notice that the solution of the KG-oscillators in (\ref%
{eq.12}) is known to be given by%
\begin{equation}
\phi \left( r\right) =C\,\exp \left( -\frac{\Omega }{2}r^{2}\right)
\,H_{C}\left( -\Omega \,\tilde{\kappa},-\frac{1}{2},0,\frac{K_{3}^{2}\tilde{%
\kappa}^{2}+K_{1}^{2}\tilde{\kappa}}{4},\frac{1-K_{2}^{2}-K_{3}^{2}\tilde{%
\kappa}^{2}-K_{1}^{2}\tilde{\kappa}}{4},\frac{r^{2}+\tilde{\kappa}}{\tilde{%
\kappa}}\right) ,  \label{eq.17}
\end{equation}%
as reported by Pereira et al. \cite{ref5} in their Eq.(11) along with the
substitution used to obtain their Eq.(8). In our opinion, nevertheless, the
form of the confluent Heun function $H_{C}$ above should never dictate the
form of the power series expansion (as the power series used by Pereira et
al. \cite{ref5} in their Eq.(12)). Instead, one should seek a power series
solution that is reducible to the KG-oscillator's one at $\tilde{\kappa}=0$
(no EiBI-gravity), $\sigma =0$ (no Wu-yang monopole), and $\alpha =1$ (flar
Minkowski spacetime). In the current methodical proposal, therefore, we
follow a different route, but a textbook one, that allows not only
reducible to the KG-oscillator's in the flat Minkowski spacetime but also
allows the so called \emph{conditional exact solvability} for the
KG-oscillators in a GM spacetime in EiBI-gravity and a WYMM. 

In so doing, we use%
\begin{equation}
\phi \left( r\right) =\exp \left( -\frac{\Omega }{2}r^{2}\right) \,R\left(
r\right) ,  \label{eq.18}
\end{equation}%
in (\ref{eq.14}), to obtain%
\begin{equation}
\left( r^{2}+\tilde{\kappa}\right) R^{\prime \prime }\left( r\right) +\left[
2\left( 1-\Omega \,\tilde{\kappa}\right) r+\frac{\tilde{\kappa}}{r}-2\Omega
r^{3}\right] R^{\prime }\left( r\right) +\left[ P_{1}r^{2}-P_{2}\right]
R\left( r\right) =0,  \label{eq.19}
\end{equation}%
where%
\begin{equation}
P_{1}=\mathcal{E}^{2}+\Omega ^{2}\tilde{\kappa}-3\Omega ,\,P_{2}=\tilde{\ell}%
\left( \tilde{\ell}+1\right) +2\Omega \,\tilde{\kappa}.  \label{eq.20}
\end{equation}%
Let us use the change of variables $y=r^{2}$ to obtain%
\begin{equation}
y\left( y+\,\tilde{\kappa}\right) R^{\prime \prime }\left( y\right) +\left[
\left( \frac{3}{2}-\Omega \,\tilde{\kappa}\right) y+\,\tilde{\kappa}-\Omega
y^{2}\right] R^{\prime }\left( y\right) +\left[ \tilde{P}_{1}y-\tilde{P}_{2}%
\right] R\left( y\right) =0;\;\tilde{P}_{i}=\frac{P_{i}}{4}.  \label{eq.21}
\end{equation}%
We may now suggest a power series solution in the form of%
\begin{equation}
R\left( y\right) =y^{\nu }\sum\limits_{j=0}^{\infty }C_{j}y^{j},
\label{eq.22}
\end{equation}%
where $\nu $ is a parameter to be determined in the process, the usefulness
of which is to be clear in the sequel. Under such settings, one obtains,
using (\ref{eq.21}),%
\begin{equation}
\sum\limits_{j=0}^{\infty }C_{j}\left\{ \left[ \tilde{P}_{1}-\Omega \left(
j+\nu \right) \right] y^{j+\nu +1}+\left[ \left( j+\nu +\frac{1}{2}-\Omega \,%
\tilde{\kappa}\right) \left( j+\nu \right) -\tilde{P}_{2}\right] y^{j+\nu }+%
\tilde{\kappa}\left( j+\nu \right) ^{2}y^{j+\nu -1}\right\} =0.
\label{eq.23}
\end{equation}%
Consequently, 
\begin{gather}
\sum\limits_{j=0}^{\infty }\left\{ C_{j}\,\left[ \tilde{P}_{1}-\Omega \left(
j+\nu \right) \right] +C_{j+1}\left[ \left( j+\nu +1\right) \left( j+\nu +%
\frac{3}{2}-\Omega \,\tilde{\kappa}\right) -\tilde{P}_{2}\right] +C_{j+2}%
\left[ \tilde{\kappa}\left( j+\nu +2\right) ^{2}\right] \right\} y^{j+\nu +1}
\notag \\
+\left\{ C_{0}\left[ \nu \left( \nu +\frac{1}{2}-\Omega \,\tilde{\kappa}%
\right) -\tilde{P}_{2}\right] +C_{1}\left[ \tilde{\kappa}\left( \nu
+1\right) ^{2}\right] \right\} y^{\nu }+C_{0}\left[ \tilde{\kappa}\nu ^{2}%
\right] y^{\nu -1}=0.  \label{eq.24}
\end{gather}%
Under such settings, we obtain the set of relations 
\begin{equation}
C_{0}\left[ \tilde{\kappa}\nu ^{2}\right] =0,\;C_{0}\left[ \nu \left( \nu +%
\frac{1}{2}-\Omega \,\tilde{\kappa}\right) -\tilde{P}_{2}\right] +C_{1}\left[
\tilde{\kappa}\left( \nu +1\right) ^{2}\right] =0,  \label{eq.24.1}
\end{equation}%
and%
\begin{equation}
C_{j+2}\left[ \tilde{\kappa}\left( j+\nu +2\right) ^{2}\right] =C_{j}\,\left[
\Omega \left( j+\nu \right) -\tilde{P}_{1}\right] +C_{j+1}\left[ \tilde{P}%
_{2}-\left( j+\nu +1\right) \left( j+\nu +\frac{3}{2}-\Omega \,\tilde{\kappa}%
\right) \right] .  \label{eq.24.2}
\end{equation}%
This set of recursion relations has to be dealt with diligently, and
case-by-case, as shall be clarified in the sequel illustrative examples.
Such illustrative examples are considered so that one obtains the exact
solution for the KG-oscillators in a GM spacetime and a WYMM in no
EiBI-gravity (i.e., $\tilde{\kappa}=0$), a conditionally exact solution for
KG-oscillators in a GM spacetime within EiBI-gravity, $\tilde{\kappa}\neq 0$,
and a WYMM, and a conditional exact solution for the massless KG-oscillators
in a GM spacetime in WYMM and Coulomb plus linear Lorentz scalar potential
within EiBI-gravity.

\subsection{KG-oscillators in a GM spacetime in a Wu-Yang magnetic monopole
in no E\lowercase{i}BI-gravity, $\tilde{\kappa}=0.$}

At this point, we may test this procedure above and remove EiBI-gravity
(i.e., $\tilde{\kappa}=0$) to reduce the problem into KG-oscillators in GM
spacetime with a WYMM. This would imply that Eq.(\ref{eq.24.1}), with $%
C_{0}\neq 0$ and $\tilde{P}_{2}=P_{2}/4$ ( $P_{2}$ is given in (\ref{eq.20}%
)), now reads%
\begin{equation}
C_{0}\left[ \nu \left( \nu +\frac{1}{2}\right) -\tilde{P}_{2}\right]
=0\Leftrightarrow \nu ^{2}+\frac{\nu }{2}-\frac{\tilde{\ell}\left( \tilde{%
\ell}+1\right) }{4}=0\Leftrightarrow \nu =\frac{\tilde{\ell}}{2},\;\nu =-%
\frac{\left( \tilde{\ell}+1\right) }{2}.  \label{eq.25}
\end{equation}%
Therefore, we shall take $\nu =\tilde{\ell}/2$ \ for it secures finiteness
of the radial wavefunction at $y=0$ (i.e., $r=0$) in (\ref{eq.22}).
Consequently, Eq. (\ref{eq.24.2}) collapses into the two-terms recursion
relation%
\begin{equation}
C_{j}\left[ \tilde{P}_{1}-\Omega \left( j+\frac{\tilde{\ell}}{2}\right) %
\right] +C_{j+1}\left[ \left( j+\frac{\tilde{\ell}}{2}+1\right) \left( j+%
\frac{\tilde{\ell}}{2}+\frac{3}{2}\right) -\tilde{P}_{2}\right]
=0,\;j=0,1,2,\cdots .  \label{eq.26}
\end{equation}%
Next, we need to truncate the power series into a polynomial of order $%
n_{r}=0,1,2,\cdots $ (to secure square integrability of the wavefunction) by
requiring that for $\forall j=n_{r}$ we have $C_{n_{r}+1}=0$ to imply%
\begin{equation}
C_{n_{r}}\left[ \tilde{P}_{1}-\Omega \left( n_{r}+\frac{\tilde{\ell}}{2}%
\right) \right] =0\Leftrightarrow \tilde{P}_{1}=\Omega \left( n_{r}+\frac{%
\tilde{\ell}}{2}\right) \Leftrightarrow \mathcal{E}^{2}=2\Omega \left(
2n_{r}+\tilde{\ell}+\frac{3}{2}\right) .  \label{eq.27}
\end{equation}%
This is the exact textbook result for the KG-oscillators in a GM spacetime
with a WYMM in no EiBI-gravity (i.e., for $\tilde{\kappa}=0$ in (\ref{eq.14}%
)). Moreover, with $\mathcal{E}^{2}$ given in (\ref{eq.15}), we one obtains%
\begin{equation}
\frac{E^{2}-m_{\circ }^{2}}{\alpha ^{2}}-3\Omega =2\Omega \left( 2n_{r}+%
\tilde{\ell}+\frac{3}{2}\right) \Leftrightarrow E=\pm \sqrt{m_{\circ
}^{2}+4\Omega \alpha ^{2}\left( n_{r}+\frac{1}{4\alpha }\sqrt{\alpha
^{2}+4\ell \left( \ell +1\right) -4\sigma ^{2}}+\frac{5}{4}\right) }.
\label{eq.28}
\end{equation}%
This result represents an exact solution for the KG-oscillators in a GM
spacetime with a WYMM in no EiBI-gravity. Moreover, it is in exact accord
with that reported in Eq. (38) of Bragan\c{c}a et al. \cite{Re11} and in
(31) of Mustafa \cite{ref017}. The coefficients of the our polynomial in (%
\ref{eq.26}), on the other hand, are now given by%
\begin{equation}
C_{j+1}=C_{j}\left[ \frac{4\Omega \left( n_{r}-j\right) }{\tilde{\ell}\left( 
\tilde{\ell}+1\right) -\left( 2j+\tilde{\ell}+2\right) \left( 2j+\tilde{\ell}%
+3\right) }\right] ;\;j=0,1,2,\cdots .  \label{eq.28.1}
\end{equation}%
Under such settings, for $j=0$ we get,%
\begin{equation}
C_{1}=C_{0}\left[ \frac{4\Omega n_{r}}{\tilde{\ell}\left( \tilde{\ell}%
+1\right) -\left( \tilde{\ell}+2\right) \left( \tilde{\ell}+3\right) }\right]
,  \label{eq.28.2}
\end{equation}%
for $j=1$ we get%
\begin{equation}
C_{2}=C_{1}\left[ \frac{4\Omega \left( n_{r}-1\right) }{\tilde{\ell}\left( 
\tilde{\ell}+1\right) -\left( \tilde{\ell}+4\right) \left( \tilde{\ell}%
+5\right) }\right] ,  \label{eq.28.3}
\end{equation}%
and so on. One should notice that we may consider $C_{0}=1$ hereinafter.
Moreover, the radial part of the wave function is therefore given by 
\begin{equation}
R\left( y\right) =y^{\tilde{\ell}/2}\sum\limits_{j=0}^{n_{r}}C_{j}y^{j}%
\Longleftrightarrow R\left( r\right) =r^{\tilde{\ell}}\sum%
\limits_{j=0}^{n_{r}}C_{j}\,r^{2j}.  \label{q.28.4}
\end{equation}

\subsection{KG-oscillators in a GM spacetime within E\lowercase{i}BI-gravity, $\tilde{\kappa}\neq 0$, and in a WYMM}

.We now consider the more general case where EiBI-gravity is involved in the
process. In this case, equation (\ref{eq.24}) suggests that in  $C_{0}\left( 
\tilde{\kappa}\nu ^{2}\right) =0\Leftrightarrow \nu =0$  since $C_{0}\neq
0\neq \tilde{\kappa}$, and%
\begin{equation}
C_{1}\tilde{\kappa}=C_{0}\tilde{P}_{2}\Longleftrightarrow C_{1}=C_{0}\frac{%
\tilde{P}_{2}}{\tilde{\kappa}}\Longleftrightarrow C_{1}=\frac{\tilde{P}_{2}}{%
\tilde{\kappa}},\;C_{0}=1.  \label{eq.29}
\end{equation}%
Moreover, equation (\ref{eq.24.2}) now reads%
\begin{equation}
C_{j+2}\left[ \tilde{\kappa}\left( j+2\right) ^{2}\right] =C_{j+1}\left[ 
\tilde{P}_{2}-\left( j+1\right) \left( j+\frac{3}{2}-\Omega \,\tilde{\kappa}%
\right) \right] +C_{j}\,\left[ \Omega \,j-\tilde{P}_{1}\right] .
\label{eq.30}
\end{equation}%
We now wish to truncate the power series into a polynomial of order $n_{r}+1$
(or if you wish a polynomial of order $n=n_{r}+1\geq 1$) so that $\forall
j=n_{r}$ we have $C_{n_r+2}=0,\, C_{n_r+1}\neq 0$, and $C_{n_r}\neq 0$.  We impose, moreover, the condition that
\begin{equation}
\tilde{P}_{2}-\left( n_r+1\right) \left( n_r+\frac{3}{2}-\Omega \,\tilde{\kappa}\right)=0. \label{eq.31}
\end{equation}%
This condition would facilitate \textit{conditional exact solvability} of the problem at hand, as well as it provides a correlation, between the KG-oscillator frequency $\Omega$ and the Eddington gravity parameter $\tilde{\kappa}$,  given by 
\begin{equation}
\tilde{P}_{2}=\left( n_{r}+1\right) \left( n_{r}+\frac{3}{2}-\Omega \,\tilde{%
\kappa}\right) \Longleftrightarrow \Omega \,\tilde{\kappa}=\frac{\alpha
^{2}\left( 4n_{r}^{2}+10n_{r}+6\right) +\sigma ^{2}-\ell \left( \ell
+1\right) }{2\alpha ^{2}\left( 2n_{r}+4\right) }.  \label{eq.32}
\end{equation}%
Consequently, since $C_{n_{r}}\neq 0$ and
\begin{equation}
 C_{n_{r}}\left[ \tilde{P%
}_{1}-\Omega \,n_{r}\right] =0\Longrightarrow \tilde{P}_{1}=\Omega \,n_{r}, \label{eq.32.01} 
\end{equation}%
and all states' energies are given by   
\begin{equation}
E=\pm \sqrt{m_{\circ }^{2}+2\alpha ^{2}\Omega \left( 2n_{r}+3\right) }=\pm 
\sqrt{m_{\circ }^{2}+\frac{\left( 2n_{r}+3\right) }{\tilde{\kappa}\left(
2n_{r}+4\right) }\left[ \alpha ^{2}\left( 4n_{r}^{2}+10n_{r}+6\right)
+\sigma ^{2}-\ell \left( \ell +1\right) \right] },  \label{eq.33}
\end{equation}%
where we have used $\Omega $ in (\ref{eq.32}). At this point, we have to emphasis that the condition used by Pereira et al. \cite{ref5} (i.e.,  non-vanishing coefficients of $C_{n_r +1}\neq 0$ in (\ref{eq.30})) have constrained the validity of their reported solution to hold true for only $n_r=0$ states. Of course, their reported solution is one of a variety of \textit{conditionally exact solutions}. Our condition in (\ref{eq.31}) is yet another feasible \textit{conditionally exact one}, therefore. Moreover, our result in (\ref{eq.32.01}) is in exact accord with that reported in Eq. (14) of Ishkhanyan et al \cite{RR1}, with proper parametric mappings, of course. 
One should observe, however, that the correlation in (\ref{eq.32}) suggests that since $\Omega >0$ and $%
\tilde{\kappa}>0$ (by definition), then the angular momentum quantum number should satisfy the relation%
\begin{equation}
\ell \left( \ell +1\right) <\alpha ^{2}\left( 4n_{r}^{2}+10n_{r}+6\right)
+\sigma ^{2}\Longleftrightarrow \ell <-\frac{1}{2}+\sqrt{\frac{1}{4}+\alpha
^{2}\left( 4n_{r}^{2}+10n_{r}+6\right) +\sigma ^{2}},  \label{eq.32.1}
\end{equation}%
where $\ell =0,1,2,\cdots $ . Hence, the maximum allowed values of $\ell $
are determined by the relation (\ref{eq.32.1}) above. For example: for  $%
\alpha =1$, $\sigma =0$, and $n_{r}=0$ we have $\ell <2$ (i.e., only states
with $\ell =0,1$ are allowed), for $\alpha =1/2$, $\sigma =0$, and $n_{r}=0$
we have $\ell <0.725$ (i.e., only $\ell =0\ $is allowed), and so on so
forth. Moreover, one should also observe that our correlation  (\ref{eq.32})
manifestly classifies our solution (\ref{eq.33}) as a \emph{conditionally
exact solution} (as known in the literature, e.g., \cite{ref8}). This
solution is not valid for $\Omega =0$ and/or $\tilde{\kappa}=0$ (note that $%
\tilde{\kappa}=0$ is a case discussed in the subsection above).  Yet, using
the left-hand-sides of Eq.s (\ref{eq.32}) and (\ref{eq.33}), our three terms
recursion relation (\ref{eq.30}) now reads%
\begin{equation}
C_{j+2}=\frac{C_{j+1}\left[ \left( n_{r}+1\right) \left( n_{r}+\frac{3}{2}%
-\Omega \,\tilde{\kappa}\right) -\left( j+1\right) \left( j+\frac{3}{2}%
-\Omega \,\tilde{\kappa}\right) \right] +C_{j}\,\Omega \,\left(
j-n_{r}\right) }{\tilde{\kappa}\left( j+2\right) ^{2}},\;j=0,1,2,\cdots ,
\label{eq.34}
\end{equation}%
yields%
\begin{equation}
C_{2}=\frac{C_{1}\left[ \left( n_{r}+1\right) \left( n_{r}+\frac{3}{2}%
-\Omega \,\tilde{\kappa}\right) -\left( \frac{3}{2}-\Omega \,\tilde{\kappa}%
\right) \right] -\,\Omega \,n_{r}}{4\tilde{\kappa}},\text{ for }j=0,
\label{eq.35}
\end{equation}%
\begin{equation}
C_{3}=\frac{C_{2}\left[ \left( n_{r}+1\right) \left( n_{r}+\frac{3}{2}%
-\Omega \,\tilde{\kappa}\right) -2\left( \frac{5}{2}-\Omega \,\tilde{\kappa}%
\right) \right] +C_{1}\,\Omega \,\left( 1-n_{r}\right) }{9\tilde{\kappa}},%
\text{ for }j=1,  \label{eq.36}
\end{equation}%
\begin{figure}[ht!]  
\centering
\includegraphics[width=0.35\textwidth]{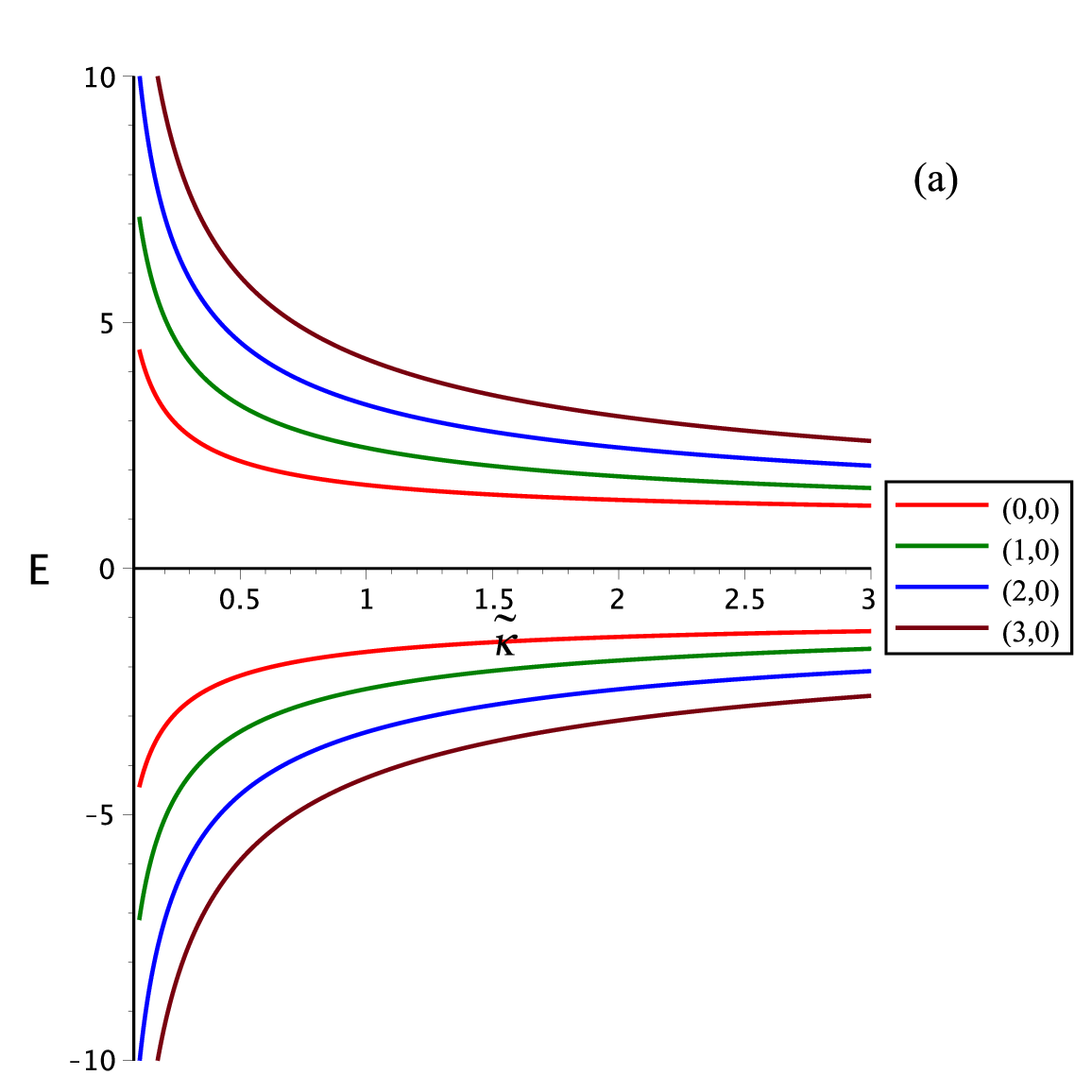}
\includegraphics[width=0.35\textwidth]{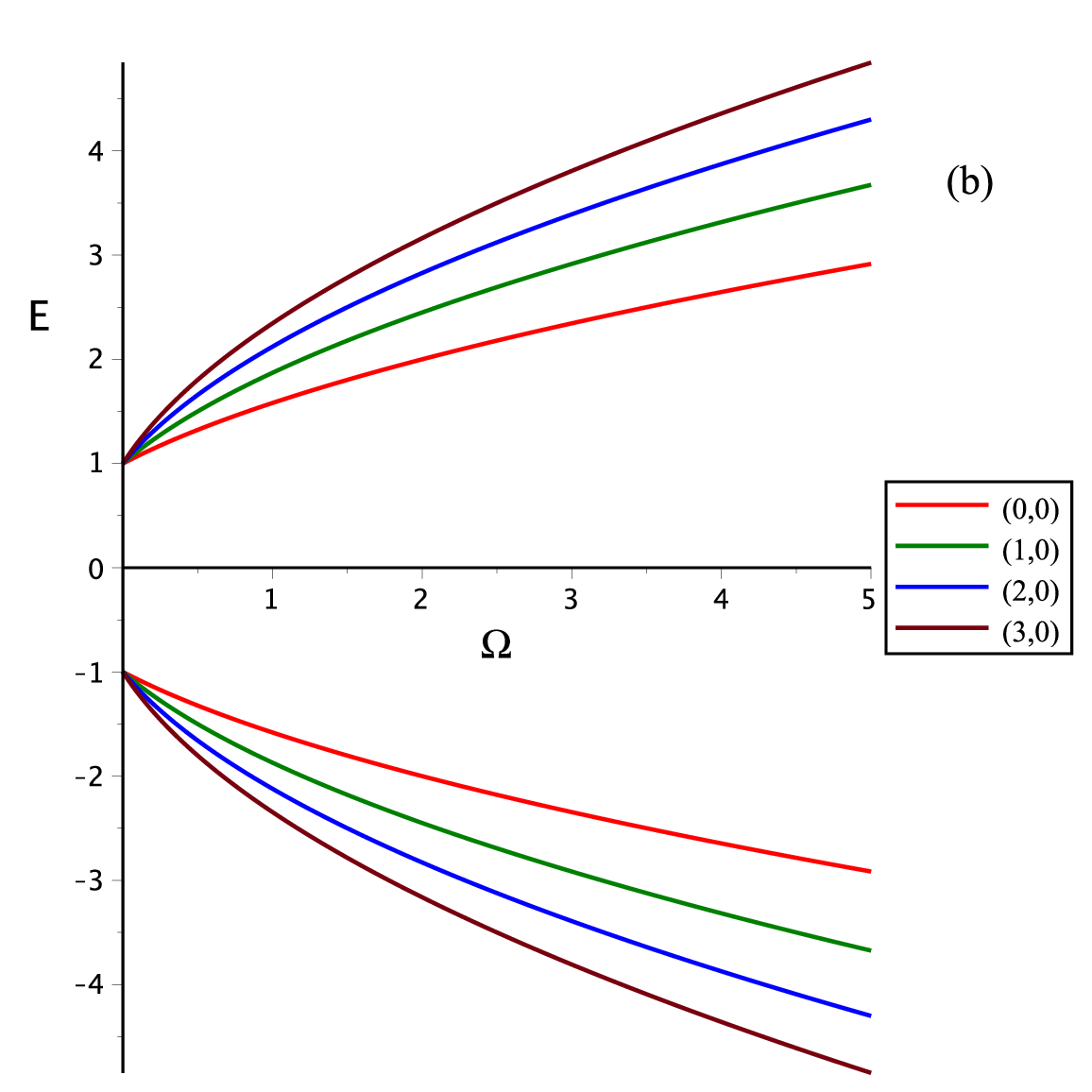}
\includegraphics[width=0.35\textwidth]{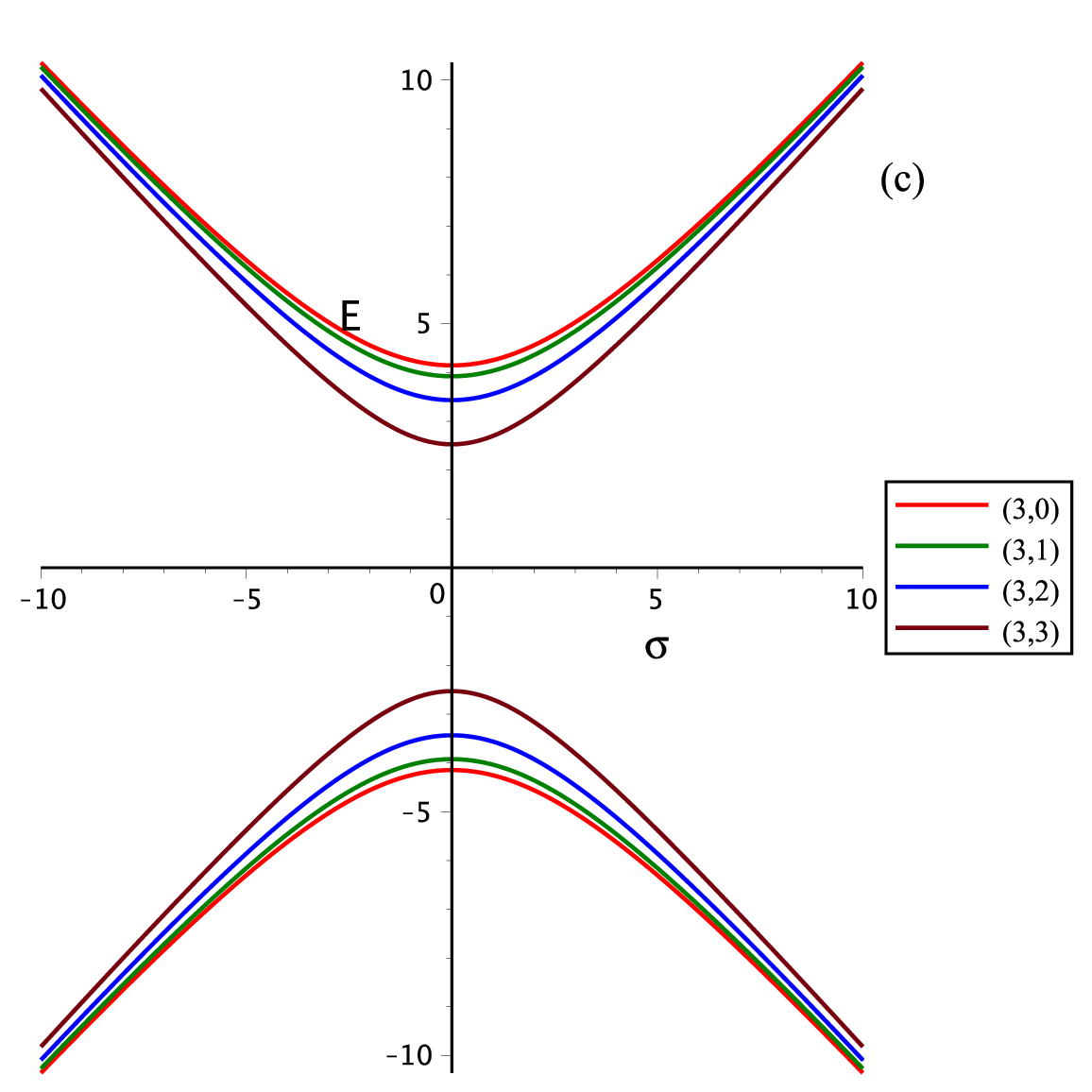}
\includegraphics[width=0.35\textwidth]{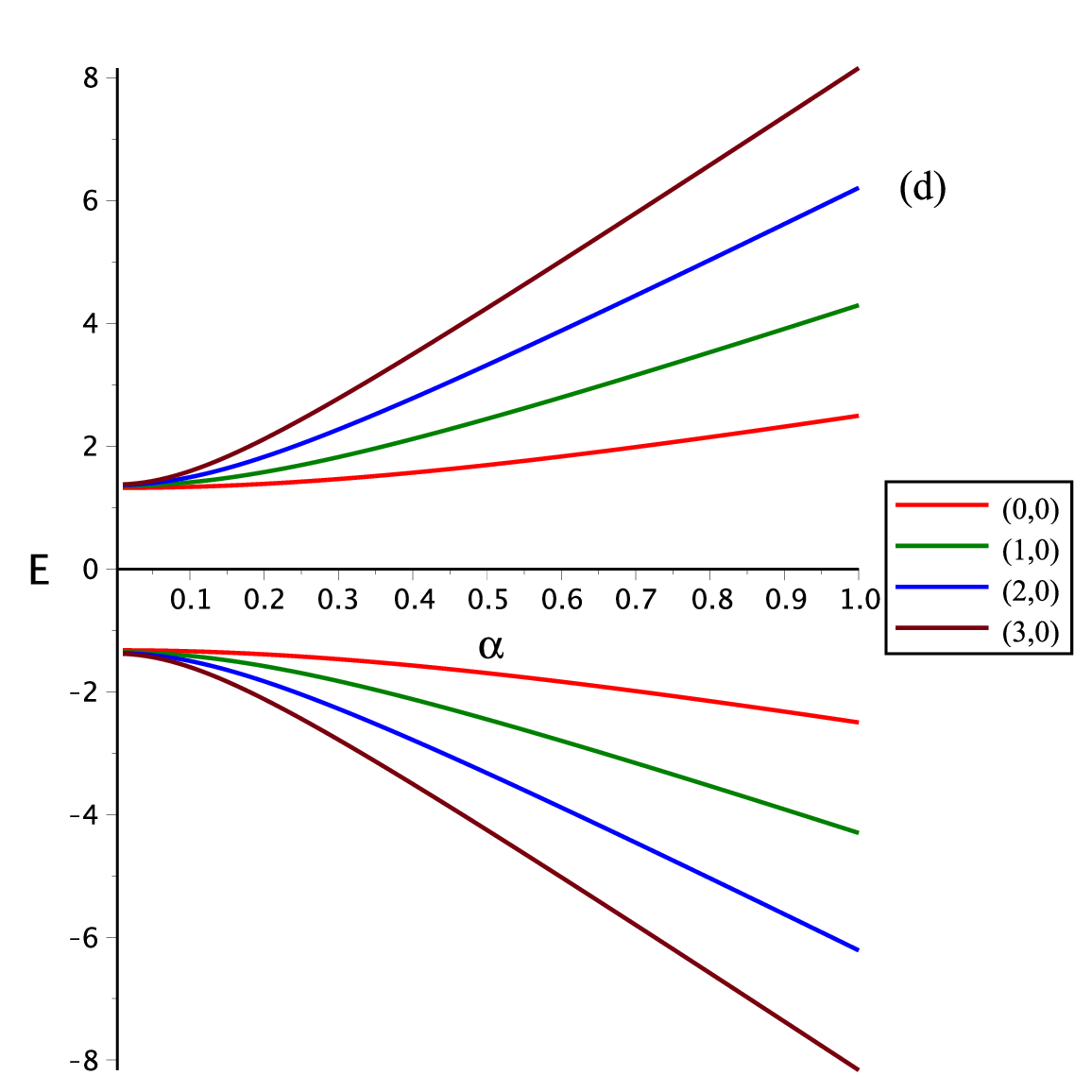}
\caption{\small 
{ The energy levels $\left( n_{r},\ell \right)$ in Eq.(\ref{eq.33}) of the KG-oscillators
in a GM spacetime within EiBI-gravity and a WYMM. The corresponding  $\left(
n_{r},\ell \right) $- states are plotted for (a) $m_{\circ }=1$, $\alpha =0.5$%
, and $\sigma =1$ for different Eddington parameter $\tilde{\kappa}>0$
values, (b) $m_{\circ }=1$, $\alpha =0.5$, and $\sigma =1$ for different of
the KG-oscillators' frequency $\Omega >0$, (c) $m_{\circ }=1$, $\alpha =0.5$%
, $\tilde{\kappa}=1$, and $\Omega =1$ for different positive and negative
values of the WYMM strength $\left\vert \sigma \right\vert $, and (d) $%
m_{\circ }=1$, $\sigma =1$, $\tilde{\kappa}=1$, and $\Omega =1$ for
different GM-parameter $0.01\leq \alpha \leq 1$ values.}}
\label{fig1}
\end{figure}%
and so on. Therefore, our power series $R\left( y\right) $ in (\ref{eq.22})
is now truncated into a polynomial of order $n=n_{r}+1\geq 1$ and is given by%
\begin{equation}
R\left( y\right) =\sum\limits_{j=0}^{n_{r}+1}C_{j}y^{j}\Leftrightarrow
R\left( r\right) =\sum\limits_{j=0}^{n_{r}+1}C_{j}\,r^{2j},  \label{eq.37}
\end{equation}%
which indeed identifies a polynomial in even powers of the $r$.

In Fig.1, we plot the energy levels in Eq.(\ref{eq.33}) taking into account
the restrictions on the orbital angular momentum quantum number Eq.(\ref%
{eq.32.1}) \ and under the conditionally exact solvability correlation (\ref%
{eq.32}).  In Fig.1(a) one may observe that the quantum states tend to
cluster for $\tilde{\kappa}>>1$ values (i.e., at extreme Eddington
gravity). Fig.1(b) shows that the separation between energy levels increase
as the oscillator frequency $\Omega $ increases. In \ Fig.1(c) energy levels
are observed to cluster at large values of the WYMM strength, i.e., $%
\left\vert \sigma \right\vert >>1$. Hereby, one should notice that we have
allowed $\sigma $ to vary from negative values to positive ones depending on
the charge involved in the WYMM structure (see Eq. (A3) in the Appendix
below). Fig.1(d), documents that the separation between the energy levels
increases with increasing GM-parameter $\alpha $. 

In connection with the results discussed above, nevertheless, the following
observation is unavoidably inviting in the process. Although the procedure
discussed above is restricted with the condition in (\ref{eq.32}), it
provides a significantly wider spectrum (for $\forall n_{r}\geq 0$, or
equivalently $\forall n=n_{r}+1\geq 1$) for the KG-oscillators in a GM
spacetime within EiBI-gravity, $\tilde{\kappa}\neq 0$, and a WYMM. The power
series approach reported by Pereira et al. \cite{ref5} cumbersome for $%
\forall n=n_{r}+1\geq 2$ states. Consequently, their solution, although\ a
conditionally exact one, it falls short and fails to address $\forall
n=n_{r}+1\geq 2$ quantum states. 

\subsection{Massless KG-oscillators in a GM spacetime within E\lowercase{i}BI-gravity
and in a WYMM under the influence of a Coulomb plus linear Lorentz scalar
potential}

We consider massless KG-oscillators in a GM spacetime in EiBI-gravity and a
WYMM subjected to a Coulomb plus linear Lorentz scalar potential $S\left(
r\right) =A/r+Br$. In this case, one may show, in a straightforward manner,
that Eq. (\ref{eq.11}) would yield%
\begin{equation}
\left\{ \left( 1+\frac{\tilde{\kappa}}{r^{2}}\right) \partial
_{r}^{2}\,+\left( \frac{2}{r}+\frac{\tilde{\kappa}}{r^{3}}\right) \partial
_{r}-\tilde{\Omega}^{2}r^{2}-\frac{\mathcal{\tilde{L}}\left( \mathcal{\tilde{%
L}+}1\right) }{r^{2}}+\mathcal{\tilde{E}}^{2}\right\} \tilde{\phi}\left(
r\right) =0,  \label{eq.38}
\end{equation}%
where,%
\begin{equation}
\tilde{\Omega}^{2}=\Omega ^{2}+\frac{B^{2}}{\alpha ^{2}},\;\mathcal{\tilde{E}%
}^{2}=\frac{E^{2}-2AB}{\alpha ^{2}}-3\tilde{\Omega}-\tilde{\Omega}^{2}\tilde{%
\kappa},\;\mathcal{\tilde{L}}\left( \mathcal{\tilde{L}+}1\right) =\frac{\ell
\left( \ell +1\right) -\sigma ^{2}+A^{2}}{\alpha ^{2}}+2\tilde{\Omega}\,%
\tilde{\kappa},  \label{eq.39}
\end{equation}%
with%
\begin{equation}
\mathcal{\tilde{L}=-}\frac{1}{2}+\sqrt{\frac{1}{4}+\frac{\ell \left( \ell
+1\right) -\sigma ^{2}+A^{2}}{\alpha ^{2}}+2\tilde{\Omega}\,\tilde{\kappa}}.
\label{eq.40}
\end{equation}%
Evidently, equation (\ref{eq.38}) is in the same form as that of Eq. (\ref%
{eq.14}). This would mandate that our equation (\ref{eq.38}) inherits the
same forms of the reported solution in the preceding subsection. That is,
with the substitution%
\begin{equation}
\tilde{\phi}\left( r\right) =\exp \left( -\frac{\tilde{\Omega}}{2}%
r^{2}\right) \,\tilde{R}\left( r\right) ,  \label{eq.41}
\end{equation}%
one obtains%
\begin{equation}
\left( r^{2}+\tilde{\kappa}\right) \tilde{R}^{\prime \prime }\left( r\right)
+\left[ 2\left( 1-\tilde{\Omega}\,\tilde{\kappa}\right) r+\frac{\tilde{\kappa%
}}{r}-2\tilde{\Omega}r^{3}\right] \tilde{R}^{\prime }\left( r\right) +\left[ 
\grave{P}_{1}r^{2}-\grave{P}_{2}\right] \tilde{R}\left( r\right) =0,
\label{eq.42}
\end{equation}%
where%
\begin{equation}
\grave{P}_{1}=\mathcal{\tilde{E}}^{2}+\tilde{\Omega}^{2}\tilde{\kappa}-3%
\tilde{\Omega},\,\grave{P}_{2}=\mathcal{\tilde{L}}\left( \mathcal{\tilde{L}+}%
1\right) +2\tilde{\Omega}\,\tilde{\kappa}.  \label{eq.43}
\end{equation}%
Let us use the change of variables $y=r^{2}$ to obtain%
\begin{equation}
y\left( y+\,\tilde{\kappa}\right) \tilde{R}^{\prime \prime }\left( y\right) +%
\left[ \left( \frac{3}{2}-\tilde{\Omega}\,\tilde{\kappa}\right) y+\,\tilde{%
\kappa}-\tilde{\Omega}y^{2}\right] \tilde{R}^{\prime }\left( y\right) +\left[
\breve{P}_{1}y-\breve{P}_{2}\right] \tilde{R}\left( y\right) =0;\;\breve{P}%
_{i}=\frac{\grave{P}_{i}}{4}.  \label{eq.44}
\end{equation}%
Following the same steps as in the preceding subsection, one obtains%
\begin{equation}
C_{1}=C_{0}\frac{\breve{P}_{2}}{\tilde{\kappa}}\Leftrightarrow C_{1}=\frac{%
\breve{P}_{2}}{\tilde{\kappa}},\;C_{0}=1,  \label{eq.45}
\end{equation}%
\begin{figure}[ht!]  
\centering
\includegraphics[width=0.3\textwidth]{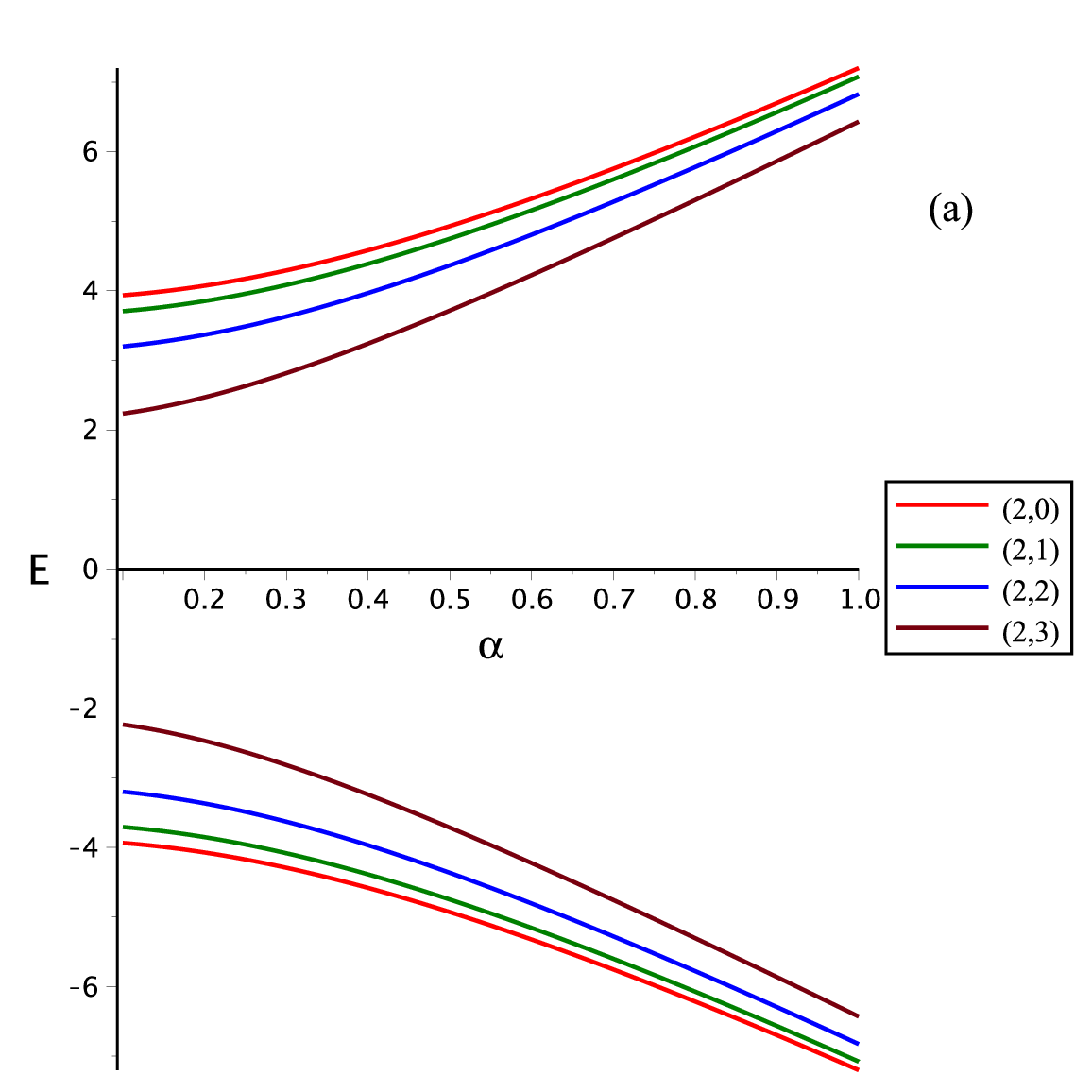}
\includegraphics[width=0.3\textwidth]{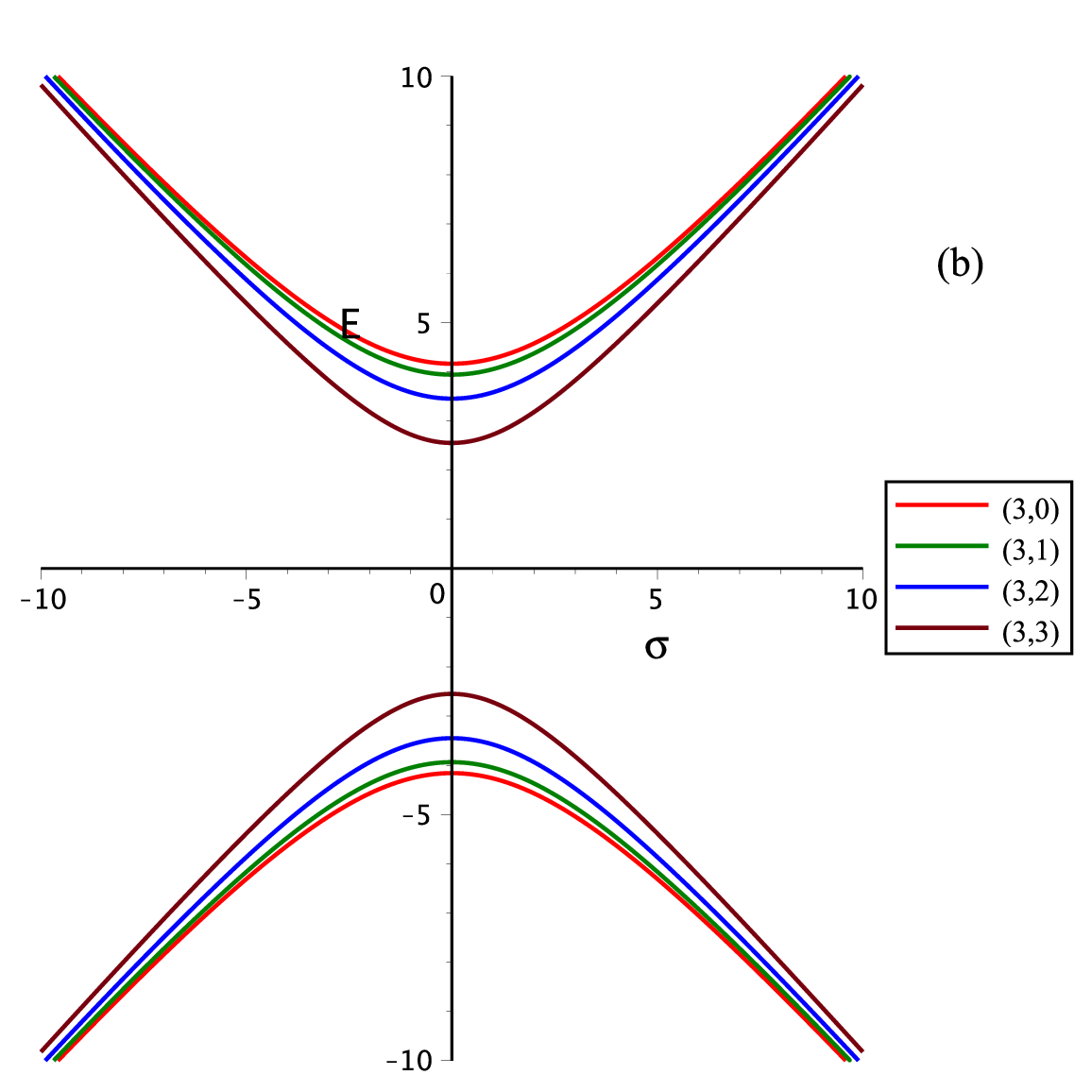}
\includegraphics[width=0.3\textwidth]{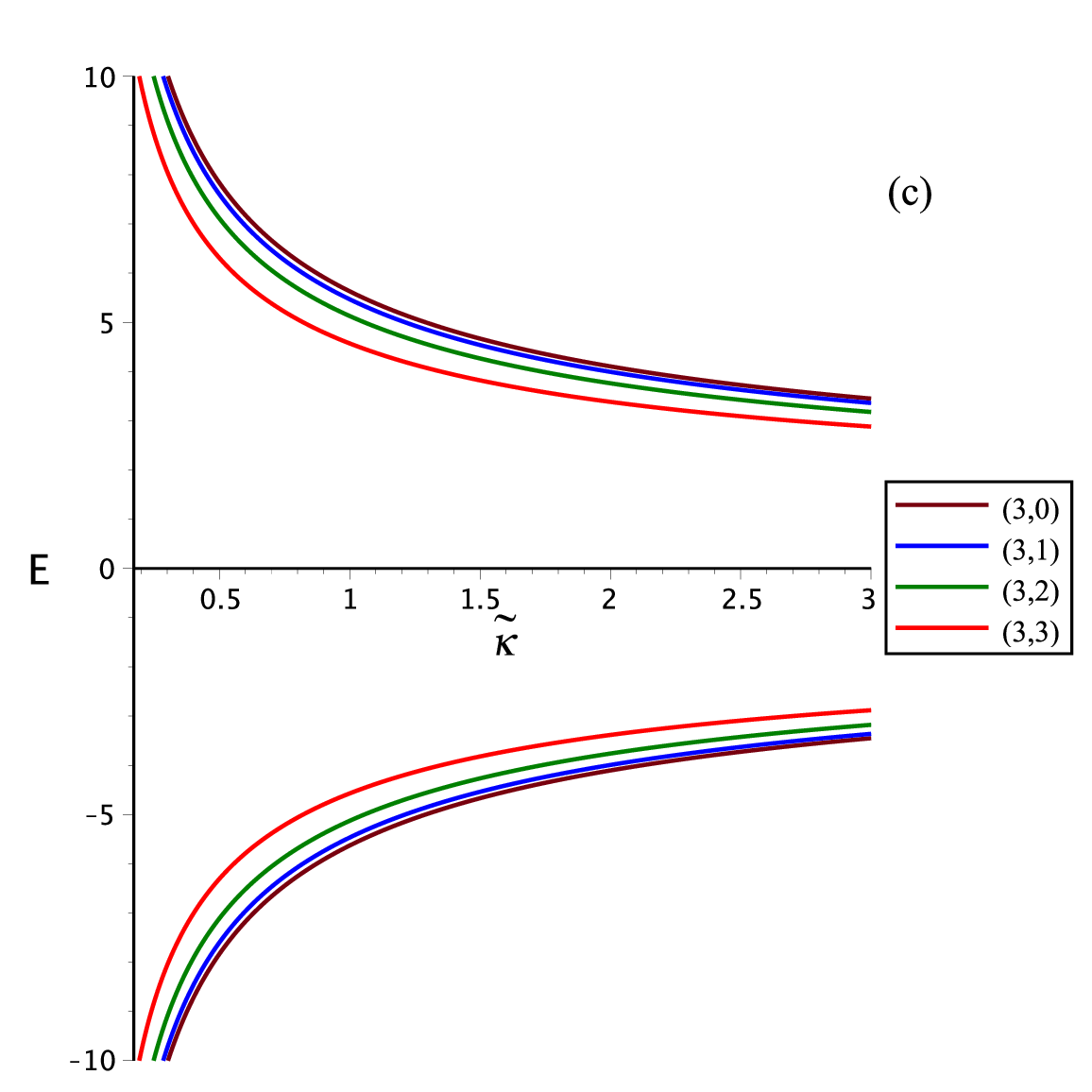}
\includegraphics[width=0.3\textwidth]{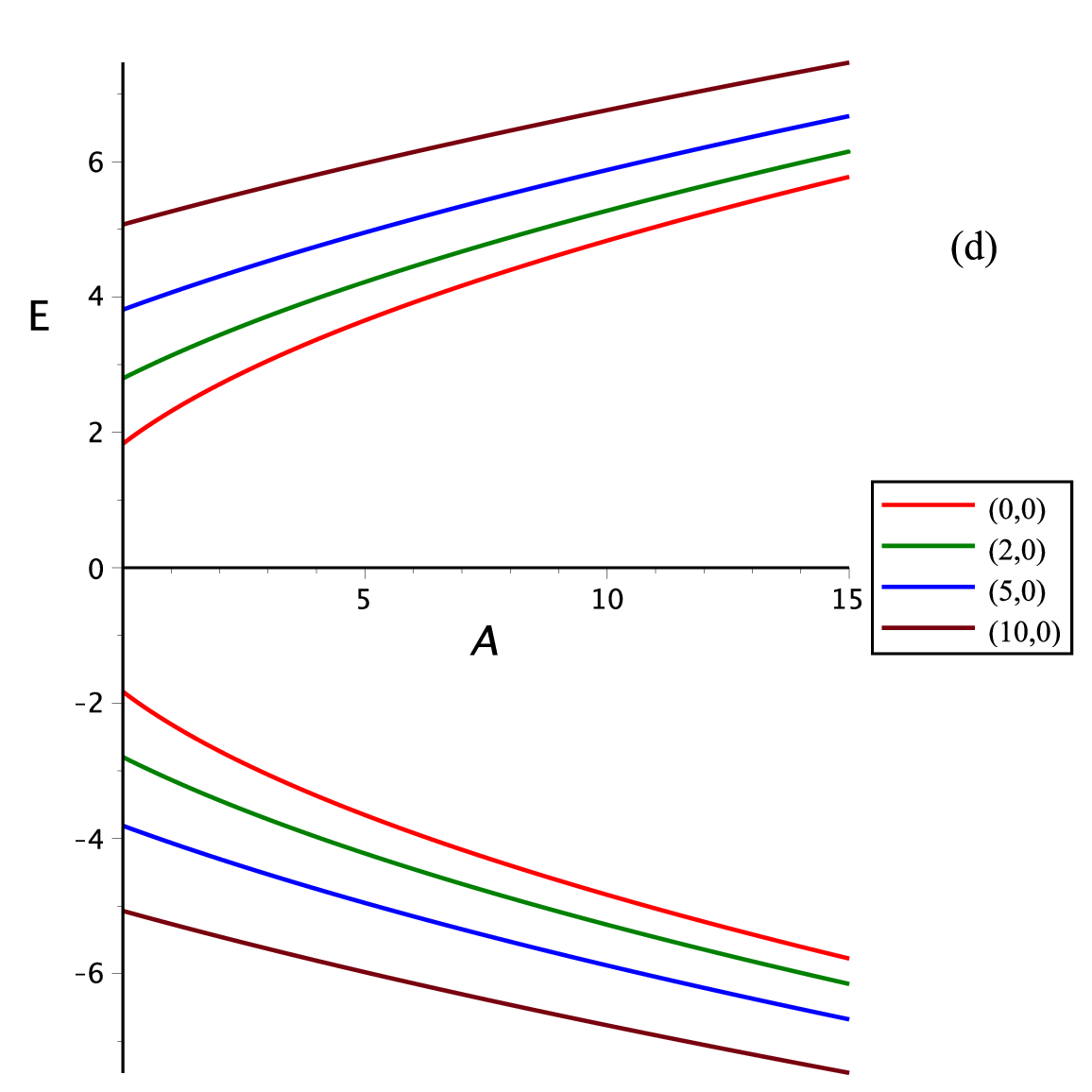}
\includegraphics[width=0.3\textwidth]{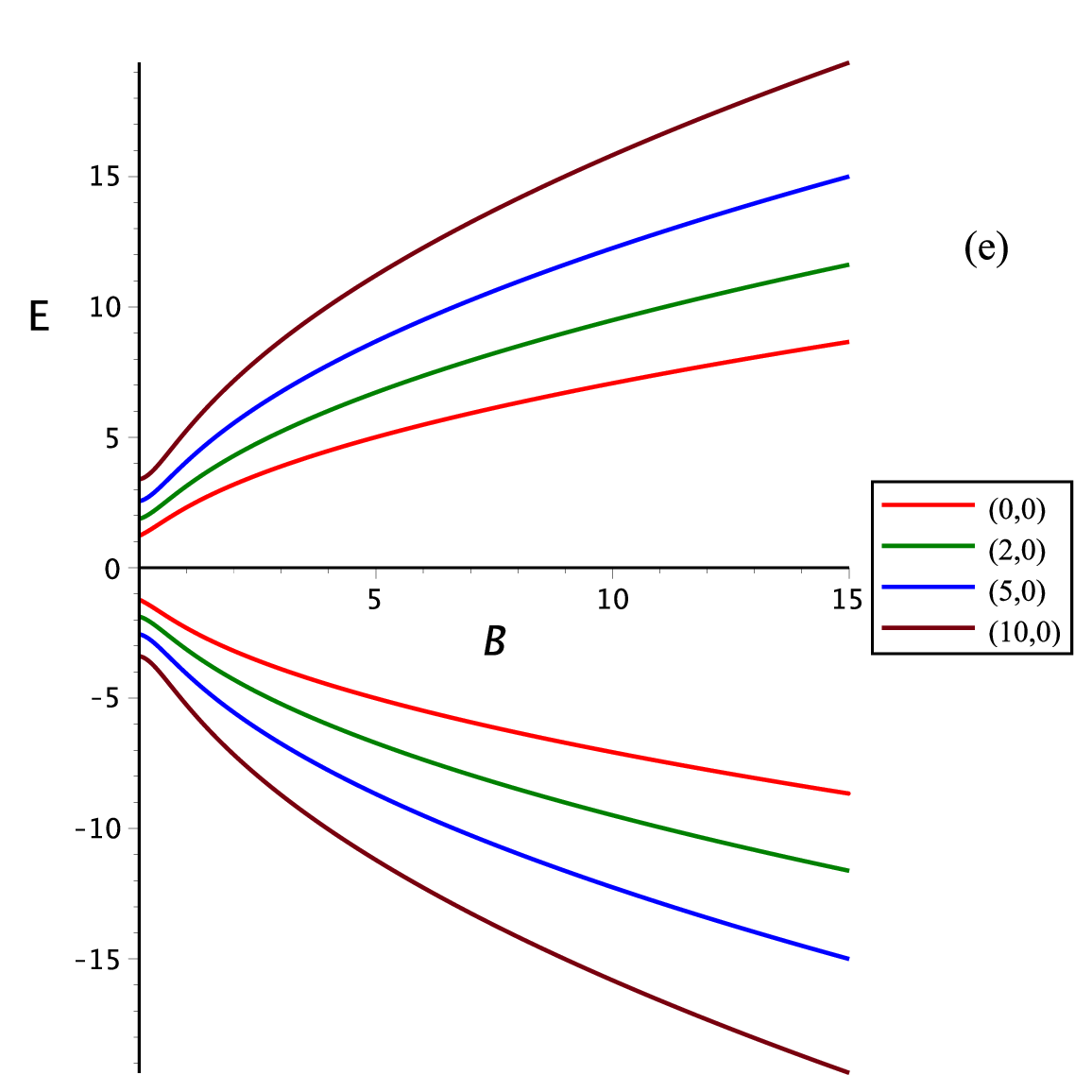}
\includegraphics[width=0.3\textwidth]{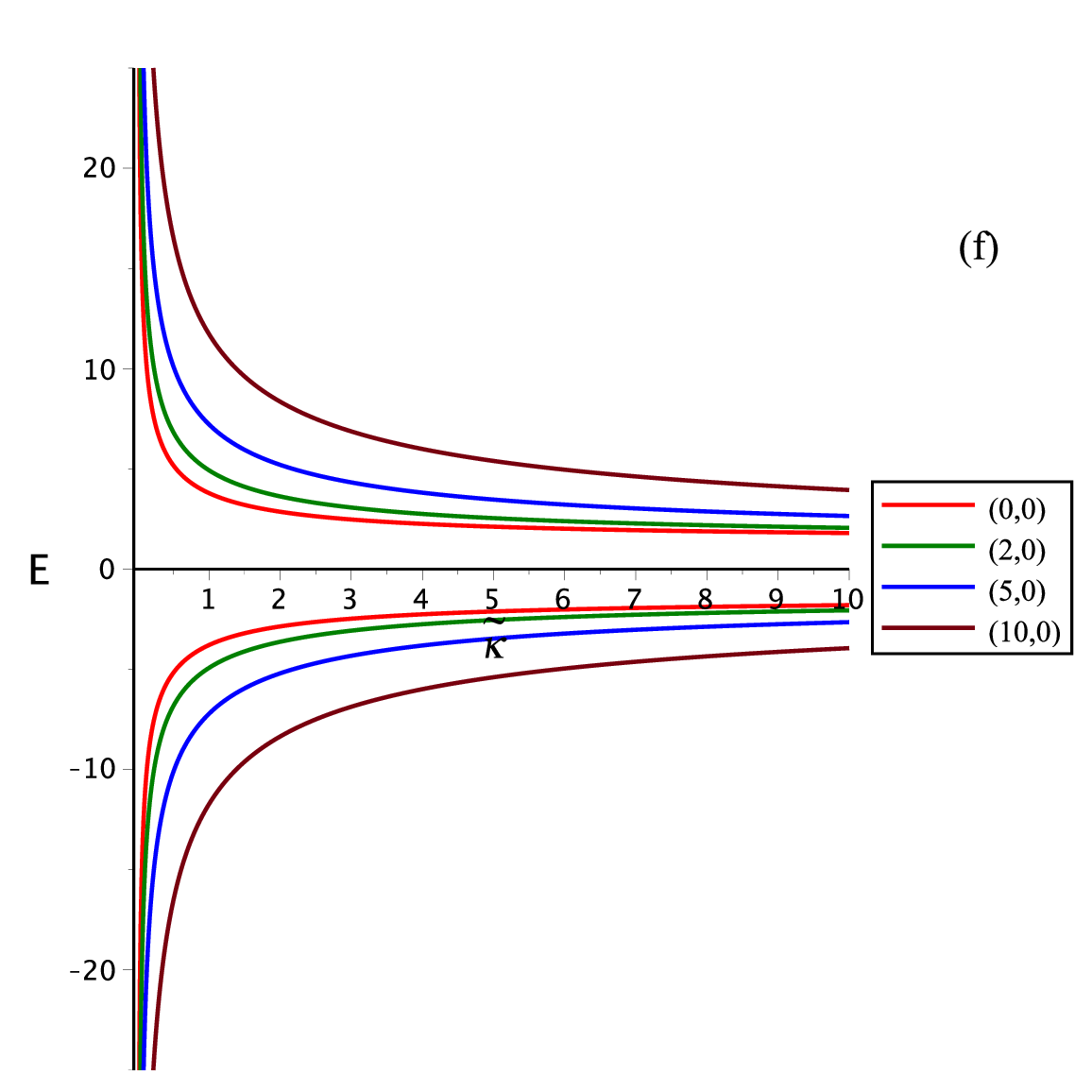}
\caption{\small 
{ The energy levels $\left( n_{r},\ell \right)$ in Eq.(\ref{eq.47}) of the massless KG-oscillators
in in a GM spacetime in EiBI-gravity and a WYMM under the influence of the
Lorentz scalar potential $S\left( r\right) =A/r+Br$. The corresponding $%
\left( n_{r},\ell \right) $- states are plotted for (a) $A=B=1$, $\sigma =4$, 
$\tilde{\kappa}=1$ for different values of $0.1\leq \alpha \leq 1$, (b) $%
A=B=1$, $\alpha =0.5$, $\tilde{\kappa}=1$ for different values of the WYMM
strength $\sigma $, (c) $A=B=1$, $\sigma =4,$ $\alpha =0.5$, $\Omega =1$ for
different values of $\tilde{\kappa}>0$, (d)  $B=1$, $\sigma =4$, $\alpha =0.5
$, $\Omega =1$ for different values of the coupling parameter $A$, (e)  $A=1$%
, $\sigma =4$, $\alpha =0.5$, $\Omega =1$ for different values of the
coupling parameter $B$, and (f)  $A=B=1$, $\sigma =1$, $\alpha =0.5$, $%
\Omega =1$ for different values of $\tilde{\kappa}>0$ and for the states
labeled $\left( n_{r},\ell \right) =\left( 0,0\right) ,\left( 2,0\right)
,\left( 5,0\right) ,\left( 10,0\right) $.}}
\label{fig2}
\end{figure}%
\begin{equation}
C_{j+2}=\frac{C_{j+1}\left[ \left( n_{r}+1\right) \left( n_{r}+\frac{3}{2}-%
\tilde{\Omega}\,\tilde{\kappa}\right) -\left( j+1\right) \left( j+\frac{3}{2}%
-\tilde{\Omega}\,\tilde{\kappa}\right) \right] +C_{j}\,\tilde{\Omega}%
\,\left( j-n_{r}\right) }{\tilde{\kappa}\left( j+2\right) ^{2}}%
,\;j=0,1,2,\cdots ,  \label{eq.46}
\end{equation}%
and%
\begin{equation}
E=\pm \sqrt{2\alpha ^{2}\tilde{\Omega}\left( 2n_{r}+3\right) +2AB},
\label{eq.47}
\end{equation}%
where the parametric correlation is obtained as%
\begin{equation}
\breve{P}_{2}=\left( n_{r}+1\right) \left( n_{r}+\frac{3}{2}-\tilde{\Omega}\,%
\tilde{\kappa}\right) \Leftrightarrow \tilde{\Omega}\,\tilde{\kappa}=\frac{%
\alpha ^{2}\left( 4n_{r}^{2}+10n_{r}+6\right) -\left[ \ell \left( \ell
+1\right) -\sigma ^{2}+A^{2}\right] }{2\alpha ^{2}\left( 2n_{r}+4\right) },
\label{eq.48}
\end{equation}%
which manifestly renders the solution to be classified as a \textit{conditionally exact solution}. This relation would again
suggest that since $\tilde{\Omega}\,>0$, and $\tilde{\kappa}>0$ then the
allowed values for angular momentum quantum number $\ell =0,1,2,\cdots $ 
are given by 
\begin{equation}
\ell \left( \ell +1\right) <\alpha ^{2}\left( 4n_{r}^{2}+10n_{r}+6\right)
+\sigma ^{2}-A^{2}\Longleftrightarrow \ell <-\frac{1}{2}+\sqrt{\frac{1}{4}%
+\alpha ^{2}\left( 4n_{r}^{2}+10n_{r}+6\right) +\sigma ^{2}-A^{2}}.
\label{eq.48.1}
\end{equation}%
Moreover, our radial part of the wave function  $\tilde{R}\left( y\right) $
is similar to that in (\ref{eq.22}) and is again a polynomial in even powers
of the $r$ given by%
\begin{equation}
\tilde{R}\left( y\right)
=\sum\limits_{j=0}^{n_{r}+1}C_{j}y^{j}\Leftrightarrow \tilde{R}\left(
r\right) =\sum\limits_{j=0}^{n_{r}+1}C_{j}\,r^{2j},  \label{eq.49}
\end{equation}

In Fig.2, we plot the energy levels in Eq.(\ref{eq.47}) along with the
restrictions on the orbital angular momentum quantum number Eq.(\ref{eq.48.1}%
) \ and under the conditionally exact solvability correlation (\ref{eq.48}).
In Fig.2(a) we observe that the separation between the energy levels
decreases with increasing GM-parameter $\alpha .$ Fig.2(b) shows that an
obvious clustering of the energy levels for $\left\vert \sigma \right\vert
>>1$. The clustering of the energy levels is also observed for $\tilde{\kappa%
}>>1$ in Fig.2(c). In Fig.2(d) we observe that the spacing between energy
levels decreases with increasing values of $A$. Whereas, in Fig.2(e) the
separation between energy levels increase with increasing values of $B$.
Finally, Fig. 2(f) shows that the effect of Eddington parameter $\tilde{%
\kappa}$ remains the same as that in Eq.(\ref{eq.33}) and shows the tendency to cluster the energy levels for $\tilde{\kappa}>>1$.
\section{Concluding remarks}
In the current proposal, we considered KG-particles in GM spacetime within EiBI-gravity and in a WYMM. We recollected/recycled the corresponding angular and radial parts of the KG-equations and brought it into a general one-dimensional radial Schr\"{o}dinger form (in section 2). We discussed a set of KG-oscillators in a GM spacetime within EiBI-gravity and including  a WYMM.  We argued that although the corresponding Schr\"{o}dinger-like KG radial equation has a
solution in the form of confluent Heun functions $H_{C}\left( a,b,c,d,f,%
\frac{r^{2}+\kappa }{\kappa }\right) $, Eq. (\ref{eq.17}), the power-series expansion should not be dictated by the argument of the confluent Heun functions (i.e., $%
\left( r^{2}+\kappa \right) /\kappa $) but rather a textbook one. That would in turn allow reduction of the three terms recursion relation, Eq. (\ref{eq.24.2} ) into that of the KG-oscillator in no Eddington gravity (i.e. $\kappa =0$). We have shown, in the same section, how to obtain the energy levels for the KG-oscillators in a GM spacetime in
a Wu-Yang magnetic monopole in no EiBI-gravity, $\tilde{\kappa}=0$, from the same recursion relations (see subsection 3-A).  Following the same recursion relations, we reported a \textit{conditionally exact solution}, through a parametric correlation, and obtain \textit{conditionally exact energy levels }for the KG-oscillators in a GM
spacetime within EiBI-gravity, $\tilde{\kappa}\neq 0$, and a WYMM (in subsection 3-B). We, moreover, discussed (in subsection 3-C) the \textit{conditionally exact} energy levels for some massless KG-oscillators in a GM spacetime within EiBI-gravity and in a WYMM under the influence of a Coulomb plus linear Lorentz scalar potential.

In connection with the energy profiles of the KG-oscillators' models considered, we observe a common effect for the Eddington parameter $\tilde{\kappa}\neq 0$. This effect is characterized by the energy levels clustering when $\kappa$  grows up from just above zero. One would, therefore, anticipate that as $\kappa >> 1$ all energy levels would cluster around $E=\pm m_\circ$ (i.e., all energy levels approach the relativistic rest mass energy gap boarders).  This is documented in Figures 1(a), 2(c), and 2(f). Whereas, the WYMM strength parameter $\sigma$ has the same effect of clustering the energy levels as $|\sigma| >>1$, but, however, the energy levels are quadratically pushed far from the rest mass energy (documented in Figures 1(c) and 2(b)).

Recently, several quantum systems described in non-trivial backgrounds have been used to analyze thermal effects, from which it is possible to describe the thermodynamic properties of these systems and the effects of the parameters that characterize the non-triviality of these backgrounds on thermodynamic quantities, such as internal energy. , entropy, specific heat, etc \cite{ter,ter1,ter2,ter3,ter4,me5,me6}. The systems presented here could, in the near future, be a gateway to analyzing the thermodynamic properties of a KG-particle in EiBI gravity.

\section{Appendix: The eigenvalues of the angular part in a Wu-Yang magnetic monopole, E{\lowercase{q}}.(\ref{eq.9}).}

In this Appendix we consider the angular part, Eq.(\ref{eq.9}), and show how
to obtain the corresponding eigenvalues in Eq.(\ref{eq.10}) for the
KG-particles in GM spacetime within EiBI-gravity and in a WYMM , discussed
in section 2. In so doing, we recall that Wu and Yang \cite{Re101,Re1011}
have introduced a magnetic monopole that is free of strings of singularities
around it (c.f., e.g., \cite{Re7,Re8,Re9,Re91,Re101,Re102}), and have
defined the vector potential $A_{\mu }$ in two regions, $R_{a}$ and $R_{b}$
covering the whole space, outside the magnet monopole, and overlap in $R_{ab}
$ so that%
\begin{equation}
\begin{tabular}{lll}
$R_{a}:0\leq \theta <\frac{\pi }{2}+\delta \medskip ,\;$ & $\;\;r>0,\;\;$ & $%
0\leq \varphi <2\pi ,$ \\ 
$R_{b}:\frac{\pi }{2}-\delta <\theta \leq \pi ,\;$ & $\;\;r>0,\;\;$ & $0\leq
\varphi <2\pi ,\medskip $ \\ 
$R_{ab}:\frac{\pi }{2}-\delta <\theta <\frac{\pi }{2}+\delta ,$ & $\;\;r>0,\;
$ & $0\leq \varphi <2\pi ,$%
\end{tabular}
\tag{A1}
\end{equation}%
where $0<\delta \leq \pi /2$. Moreover, the 4-vector potential has a
non-vanishing component in each region given by%
\begin{equation}
A_{\varphi }=sg-g\cos \theta =\left\{ 
\begin{tabular}{ll}
$A_{\varphi ,A}=g\left( 1-\cos \theta \right) ,$ & for $s=1\medskip $ \\ 
$A_{\varphi ,B}=-g\left( 1+\cos \theta \right) ,$ & for $s=-1$%
\end{tabular}%
\right. ,  \tag{A2}
\end{equation}%
where, $A_{\varphi ,a}$ and $A_{\varphi ,b}$ are correlated by the gauge
transformation \cite{Re8,Re102} 
\begin{equation}
A_{\varphi ,a}=A_{\varphi ,b}+\frac{i}{e}S\,\partial _{\varphi
}\,S^{-1}\,;\;S=e^{2i\sigma \varphi },\;\sigma =eg.  \tag{A3}
\end{equation}%
Under such settings, the substitution of%
\begin{equation}
Y_{\sigma \ell m}\left( \theta ,\varphi \right) =e^{i\left( m+s\sigma
\right) }\ \Theta _{\sigma \ell m}\left( \theta \right) ,  \tag{A6}
\end{equation}%
in (\ref{eq.9}), where $m=0,\pm 1,\pm 2,\cdots $ is the magnetic quantum
number, would yield%
\begin{equation}
\left\{ \frac{1}{\sin \theta }\partial _{\theta }\sin \theta \;\partial
_{\theta }-\frac{\left( m+\sigma \cos \theta \right) ^{2}}{\sin ^{2}\theta }%
\right\} \Theta _{\sigma \ell m}\left( \theta \right) =-\lambda \Theta
_{\sigma \ell m}\left( \theta \right) .  \tag{A7}
\end{equation}%
At this point, one should notice that Eq.(A7) is $s$-independent so that $%
\Theta _{\sigma \ell m}\left( \theta \right) =\left[ \Theta _{\sigma \ell
m}\left( \theta \right) \right] _{a}=\left[ \Theta _{\sigma \ell m}\left(
\theta \right) \right] _{b}$ as readily observed by Wu-Yang \cite{Re101}.
Next, we use $x=\cos \theta $ and re-write Eq.(A7) as%
\begin{equation}
\left\{ \left( 1-x^{2}\right) \,\partial _{x}^{2}-2x\,\partial _{x}-\frac{%
\left( m+\sigma \,x\right) ^{2}}{1-x^{2}}\right\} \Theta _{\sigma \ell
m}\left( \theta \right) =-\lambda \Theta _{\sigma \ell m}\left( \theta
\right) .  \tag{A8}
\end{equation}%
Which, with the substitution%
\begin{equation}
\Theta _{\sigma \ell m}\left( \theta \right) =\left( 1-x\right) ^{\beta
/2}\left( 1+x\right) ^{\gamma /2}\,P_{\sigma \ell m}\left( x\right)
\,;\;\beta =|m|+\sigma ,\;\gamma =|m|-\sigma ,  \tag{A9}
\end{equation}%
would yield that%
\begin{equation}
\,P_{\sigma \ell m}\left( x\right) =C\;M\left( \frac{1}{2}+|m|\pm \sqrt{%
\sigma ^{2}+\lambda +\frac{1}{4}},|m|-\sigma +1,\frac{1}{2}\left( 1+x\right)
\right) .  \tag{A10}
\end{equation}%
However, we are interested in a finite and bounded solution for quantum
states. This would manifestly require that the hypergeometric series $%
M\left( a,b,z\right) $ be truncated into a polynomial of order $n\geq 0$
using the condition that%
\begin{equation}
\frac{1}{2}+|m|\pm \sqrt{\sigma ^{2}+\lambda +\frac{1}{4}}%
=-n\Longleftrightarrow \sigma ^{2}+\lambda =\left( n+|m|\right) \left(
n+|m|+1\right) ;\;n,|m|\,\geq 0.  \tag{A11}
\end{equation}%
One should reminded that when $\sigma =0$ (i.e., no Wu-Yang monopole effect)
our $\lambda =\ell \left( \ell +1\right) $, where $\ell =0,1,2,\cdots $ is
the angular momentum quantum number. Hence, without any lose of generality
of the solution, one may safely consider that $\ell =n+|m|$ and take 
\begin{equation}
\lambda =\ell \left( \ell +1\right) -\sigma ^{2}.  \label{A11.1}
\end{equation}%
This is exactly the result given in Eq.(\ref{eq.10}). Moreover, the so
called Wu and Yang \cite{Re8,Re101} monopole harmonics $Y_{\sigma \ell
m}\left( \theta ,\varphi \right) $ are now given by%
\begin{equation}
Y_{\sigma \ell m}\left( \theta ,\varphi \right) =\left\{ 
\begin{tabular}{ll}
$e^{i\left( m+\sigma \right) \varphi }\,\left( 1-x\right) ^{\beta /2}\left(
1+x\right) ^{\gamma /2}\,P_{\sigma \ell m}\left( x\right) ;\;\medskip $ & in
region $R_{a}$ \\ 
$e^{i\left( m-\sigma \right) \varphi }\,\left( 1-x\right) ^{\beta /2}\left(
1+x\right) ^{\gamma /2}\,P_{\sigma \ell m}\left( x\right) ;\medskip $ & in
region $R_{b}$%
\end{tabular}%
\right. .  \tag{A13}
\end{equation}%
This Appendix section makes the current methodical proposal self-contained.

\end{document}